\documentclass[12pt]{article}
\usepackage{latexsym}

\csname @addtoreset\endcsname{equation}{section}

\def\bseq{\begin{subequation}}  
\def\eseq{\end{subequation}}
\def\bsea{\begin{subeqnarray}}  
\def\esea{\end{subeqnarray}}

\newcommand{\bbox}{\lower.2ex\hbox{$\Box$}}
\newcommand{\eqn}[1]{(\ref{#1})}
\newcommand{\ft}[2]{{\textstyle\frac{#1}{#2}}}
\newcommand{\ff}[2]{$\stackrel{\textstyle #1}{\textstyle #2}$}
\newcommand{\ffm}[2]{$\pmatrix{\textstyle #1 \cr \textstyle #2}$}
\newcommand{\ghe}[3]{$\stackrel{\textstyle #1}{\scriptstyle (#2,#3)}$}
\newcommand{\lre}{\multicolumn{3}{c}{$\stackrel{*}{\longleftrightarrow}$}}

\newcommand{\beq}{\begin{equation}}
\newcommand{\eeq}{\end{equation}}
\newcommand{\bea}{\begin{eqnarray}}
\newcommand{\eea}{\end{eqnarray}}
\newcommand{\ena}{\end{eqnarray}}

\renewcommand{\a}{\alpha}
\renewcommand{\b}{\beta}

\renewcommand{\d}{\delta}
\renewcommand{\th}{\theta}

\newcommand{\pa}{\partial}

\newcommand{\g}{\gamma}

\newcommand{\m}{\mu}

\newcommand{\n}{\nu}

\newcommand{\s}{\sigma}
\renewcommand{\S}{\Sigma}

\newcommand{\Db}{\bar{D}}

\newcommand{\sigmab}{\bar{\sigma}}
\newcommand{\Sigmab}{\bar{\Sigma}}
\newcommand{\Phib}{\bar{\Phi}}

\newcommand{\ad}{{\dot{\alpha}}}

\begin{document}

\begin{titlepage}
\begin{flushright}BRX TH-406 \\ KUL-TF-97/08 \\ IFUM-553-FT\\ hep-th/9703081
\end{flushright}
\vfill
\begin{center}
{\LARGE\bf Quantization of the complex linear superfield}    \\
\vskip 27.mm  \large
{\bf   Marc Grisaru $^{1,*}$,
Antoine Van Proeyen $^{2,\dagger}$ \\ and  Daniela Zanon$^3$ } \\
\vfill
{\small
$^1$ Physics Department, Brandeis University, Waltham, MA 02254, USA \\
\vspace{6pt}
$^2$ Instituut voor theoretische fysica,
Katholieke Universiteit Leuven,\\ B-3001 Leuven, Belgium\\
\vspace{6pt}
 $^3$ Dipartimento di Fisica dell'Universit\`a di Milano and\\
INFN, Sezione di Milano, via Celoria 16,
I-20133 Milano, Italy. }
\end{center}
\vfill

\begin{center}
{\bf ABSTRACT}
\end{center}
\begin{quote}
The quantization of the complex linear superfield
requires an infinite tower of ghosts.
We use the Batalin-Vilkovisky method to obtain a gauge-fixed action.
In superspace, the method brings in some novel features
\vfill      \hrule width 5.cm
\vskip 2.mm
{\small
\noindent $^*$ Supported in part by NSF Grant No. PHY-92-22318 \\
\noindent $^\dagger$ Onderzoeksdirecteur FWO, Belgium }
\end{quote}
\begin{flushleft}
March 1997
\end{flushleft}
\end{titlepage}

\section{Introduction}
Whereas the conventional superspace description of the four-dimensional $N=1$
scalar multiplet
is in terms of a chiral scalar superfield $\Phi$ satisfying $\Db_\ad \Phi =0$
with kinetic lagrangian $\Phib \Phi$, an alternative description is by means
of a complex linear superfield $\S$ satisfying $\Db^2 \S =0$ with lagrangian
$\Sigmab \S$, the so-called nonminimal scalar multiplet \cite{superspace}.
The equivalence between the two descriptions
can be exhibited by a duality transformation, or by
examining the corresponding component actions; their auxiliary field
content is different, but the dynamical scalar and spinor degrees of freedom are
represented in the same way.

The complex linear superfield appears in various contexts in the superspace
description of supersymmetric systems. It is present as a compensator
in the nonminimal, $n \neq -\frac{1}{3} , 0$ formulation of supergravity
\cite{superspace}. It has also been
used in models describing the supersymmetric extension of the low-energy
QCD action, as a substitute for the chiral scalar superfield \cite{WZNW}.
Finally, it appears as a ghost, if one quantizes the chiral
superfield by first solving the chirality constraint in terms of a
general superfield, $\Phi = \Db^2 \Psi$.

Whereas at the classical level the complex linear superfield
presents no problems, its quantization
runs into some inherent obstacles because the constraint $\Db^2 \S =0$ is
difficult
to handle, in contrast to the familiar chirality constraint $\Db_\ad \Phi =0$.
It is not clear how to define functional
differentiation or integration with respect to $\S$:
the chiral superfield functional differentiation formula $\d \Phi (z) / \d \Phi
 (z') = \Db^2 \d^{(8)} (z-z')$ with $z \equiv (x, \th   )$ has no counterpart for
$\d \S (z) / \d \S (z')$; similarly the procedure defining chiral superfield
functional integration (see ref. \cite{superspace} sec. 3.8) has no obvious
extension to the linear superfield case.  Alternatively, if one solves the
 linearity
constraint by $\S = \Db_\ad \sigmab^\ad$,   $\Sigmab = D_\a \sigma^\a$
 in terms of unconstrained
spinor superfields $\s^\a$, $\sigmab^\ad$, functional integration over
$\s^\a$, $\sigmab^\ad$ requires gauge fixing and the
quantization leads to an infinite tower of ghosts. 
(We should emphasize that the situation is quite different for the {\em real}
linear superfield $G = \bar{G}$ satisfying $D^2G = \Db^2G=0$ which can be
presented as the field strength of a chiral spinor superfield (tensor multiplet)
$G= D_\a \phi^\a +\Db_\ad \bar{\phi}^\ad$ whose quantization presents no
difficulties, see \cite{superspace}, sec.6.2.c.)

Although at the classical level the linear superfield describes the
same physical degrees of freedom as the chiral superfield, its
quantization is not purely academic. For example, in the
quantization of nonminimal supergravity, it cannot be avoided. If
one considers, for example, quantum effects in $N=2$ Yang-Mills
coupled to $N=2$ supergravity (whose $N=1$ description necessarily
involves nonminimal, $n=-1$, supergravity) loop effects due to the
linear superfield are present \cite{YMSG}. Furthermore, since the
chiral scalar and linear scalar superfields are related by a duality
transformation at the classical level, its quantization could lead
to further understanding of how duality is affected by quantum
effects.

This paper is devoted to the quantization of the complex linear superfield,
viewed as the field strength of the unconstrained gauge superfields $\s^\a$,
$ \sigmab^\ad $. As mentioned above, one encounters the problem of
ghosts for ghosts and one unavoidably runs into an infinite tower of
ghosts whose systematic handling is best done by using the
Batalin-Vilkovisky (BV) formalism \cite{BV}. Although
the final result is difficult to use in applications, we believe it
is worthwhile to have a complete solution to the quantization
problem. Furthermore, the system under consideration
presents a nice illustration of the power of the
Batalin-Vilkovisky method in handling an intricate quantization problem.
We believe that this application of the BV formalism to superspace
quantization is new and has novel features not encountered in
ordinary space-time quantization.

Our paper is organized as follows: in section 2 we give a brief
description of the classical linear superfield and its component
decomposition, and describe qualitatively its quantization. In
section 3 we review the BV formalism. In section 4 we begin the
quantization procedure by describing the first level gauge fixing.
Section 5 presents the full result, while section 6 describes the
techniques we have used. Section 7 discusses our results. The
Appendices contain various additional tools and ancillary material.

\section{The classical theory and its gauge structure}

The kinetic action for a complex linear superfield $\S$, $\Sigmab$, with
$\Db^2 \S = D^2 \Sigmab =0$ is
\beq
S=- \int d^4x d^4 \th ~ \Sigmab \Sigma
\eeq
The components of $\S$ are given by
\bea
&& B= \S | ~~~~,~~~~ \rho_\a = D_\a \S | ~~~~,~~~~ \bar{\zeta}_\ad = \Db_\ad \S |
\nonumber\\
&&H= D^2 \S |~~~~,~~~~ p_{\a \ad} =  \Db_\ad D_\a \S |~~~~,~~~~
\bar{\b}_\ad =\frac{1}{2} D^\a \Db_\ad D_\a \S | \label{components}
\eea
and their complex conjugates.
The component action is \cite{Deo,superspace}
\beq
  S= \int d^4x [ \bar{B} \Box B - \bar{\zeta}^\ad i \pa_{\a \ad} \zeta^\a
-\bar{H} H
 +\beta^\a \rho_\a +\bar{\rho}^\ad \bar{\b}_\ad - \bar p^{\a \ad}
p_{\a \ad}]
\eeq
with propagating complex scalar and spinor degrees of freedom just like
for the standard scalar multiplet, but with a different auxiliary field
structure. (The minus sign in front of the superspace action was chosen so
that the component scalar field has the correct sign for its kinetic term.)

The equivalence of the descriptions  of the scalar multiplet by the
linear superfield $\S$ and by the chiral superfield $\Phi$ can be exhibited by
means of a duality transformation, starting with the action
\beq
S_D = -\int d^4x d^4 \th ~ [ \bar{\S} \S +\Phi \S +\Phib \bar{\S} ]
\eeq
(with { \em unconstrained} $\S$ and chiral $\Phi$) \cite{superspace}.
Using the equations of motion
 to eliminate the superfields $\S, \Sigmab$, leads to the usual chiral superfield
action. Eliminating instead the superfields  $\Phi$, $\Phib$
(whose equations of motion
impose the linearity constraint $\Db^2 \S = D^2 \Sigmab =0 $) leads
to the linear superfield action.

The linearity constraint can be solved in terms of an unconstrained
spinor superfield and its complex conjugate by
\beq
\S = \Db _\ad \sigmab^\ad ~~~~,~~~~ \Sigmab = D_\a \s^\a \ ,
\eeq
and the action becomes
\beq
S_{cl}= -\int d^4x d^4 \th ~~D_\a \s^\a \cdot \Db_\ad \sigmab^\ad
=-\int d^4x d^4 \th  ~\s^\a D_\a \Db_\ad \sigmab^\ad\ .
\label{Scl}
\eeq
The solution of the constraint has introduced some gauge invariance,
since clearly  the general spinor superfield $\s^\a$ has more components than
$\S$;
the operator $D_\a \Db_\ad$ is not invertible.
We view $\S$ as the field strength of the gauge field $\s^\a$.

The
above action is invariant under the variation $\d \s^\a = D_\b \s^{(\a \b)}$
with unconstrained {\em symmetric} (as indicated by the brackets)
bispinor gauge parameter, since
$D_\a D_\b$ is antisymmetric in the indices.
However, this variation has zero modes, $\d \s^{(\a \b)} =
D_\g \s^{(\a \b \g)}$ with the new parameter symmetric in its indices.
Similarly, we find zero modes $\d \s^{(\a \b \g)} = D_\d \s^{(\a \b \g \d)}$,
and so on. Proceeding in this manner one discovers an infinite chain of
transformations with zero modes which, upon quantization, leads to an
infinite tower of ghosts. This  comes about because at every step the ghosts have
more
components than are necessary to remove gauge degrees of freedom, an apparently
unavoidable situation if one wants to maintain manifest Lorentz invariance.
It is this feature which makes the quantization of the complex linear superfield
difficult.

 This superspace situation is somewhat
analogous to that of a bosonic theory for a vector field $V^\m$ with
lagrangian
$(\pa_\m V^\m )^2$ and gauge invariance $\d V^\m = \pa_\n V^{[\m \n]}$
in terms of a second rank {\em antisymmetric} tensor gauge parameter.
 Again one has zero modes $\pa_{\rho}V ^{[\m \n \rho] }$, etc., because the gauge
parameter has too many components.
However in this case, at least in $4$ dimensions, the tower of ghosts
ends with the fourth rank antisymmetric tensor $V^{[ \m \n \rho \s]}$.

\section{Brief account of the BV formalism}

We review in this section the Batalin-Vilkovisky quantization
\cite{BV}, with emphasis
on the case where the gauge transformation of the original gauge field
has zero modes, requiring second, and possibly further generations of ghosts.
One assigns ghost number $g$ to the $g$'th generation ghost, with the
physical field having $g=0$.
Generically we denote the physical and ghost fields
by $\Phi^A$. For each field one introduces an antifield $\Phi^*_A$
of opposite statistics, and for
functionals $F( \Phi , \Phi^*)$, $G(\Phi , \Phi^*)$ one introduces an
antibracket
\beq
(F,G) = F \frac{ \stackrel{\leftarrow}{\d}}{\d \Phi^A} \cdot
\frac{ \stackrel{\rightarrow} {\d}}{\d \Phi_A^* }G
- F \frac{ \stackrel{\leftarrow}{\d}}{\d \Phi_A^*} \cdot
\frac{ \stackrel{\rightarrow} {\d}}{\d \Phi^A }G
\eeq
(summed over $A$). One assigns ghost number $g( \Phi_A^*) = -g(\Phi^A)-1$.

One defines {\em the minimal extended action}
 $S_{min}$ by
\begin{equation}
S_{min}  =S_{cl}+ \Phi_A^* \d \Phi^A  \ ,   \label{Smin}
\end{equation}
which contains, apart from the classical action, terms involving
the transformations of all the fields,
with the gauge parameters replaced by the corresponding ghosts.
The minimal extended action has ghost number zero, and satisfies the master
equation $(S_{min}, S_{min})=0$. Moreover it has to satisfy the
`properness condition'. That is essentially the statement that for
each gauge invariance (or zero mode of gauge invariances) one has
introduced a ghost, (or a ghost for ghosts).

Gauge fixing is performed by canonical transformations on the set of
fields and antifields \cite{BVcan,bvshortrev}. Such canonical
transformations from the set $\{\Phi^A, \, \Phi^*_A\}$ to a new
basis $\{\tilde\Phi^A, \, \tilde\Phi^*_A\}$ can be determined by a
generating fermion $F(\Phi, \tilde\Phi^*)$ of ghost number $-1$, through
\begin{equation}
\tilde\Phi^A=\frac{\delta F(\Phi , \tilde\Phi^*)
}{\delta \tilde\Phi^*_A}\hspace{2cm}
\Phi^*_A=\frac{\delta F(\Phi , \tilde\Phi^*)}{\delta \Phi ^A}\ .
\label{Fcan}
\end{equation}
Generally, we use generating fermions of the form
\begin{equation}
F(\Phi , \tilde\Phi^*) =\Phi^A \tilde\Phi^*_A +\Psi(\Phi)\ ;
\end{equation}
$\Psi$, also of ghost number $-1$, is called the {\em gauge
fermion}.

So far we had not yet introduced fields with negative
ghost numbers. This is the first place where we see the necessity of
adding {\em nonminimal} fields (and their antifields). The usual
antighost is the first of these. They should be introduced in the
action without changing its physical content, and maintaining the
master equation. For example, one introduces
an antighost $b$, with ghost number $-1$ (with
its antifield $b^*$ of ghost number 0), and adds
the term $(b^*)^2$ to the extended action. In this basis $b$ itself does
not yet appear in the action, but it will after the canonical
transformation. There are several reasons
for introducing non-minimal fields, as we
will explain below. One may also be
led to introducing Nielsen-Kallosh ghosts \cite{NKgh},
as we will explain in section~\ref{ss:NK}.

In the simplest case, when there are no zero-modes, one introduces
for each gauge invariance one such field $b^i$ as nonminimal
field, and gauge-fixing functions $F_i(\sigma)$, where $\sigma$
stands here for the classical fields (of ghost number~0). The gauge
fermion is then
\beq
\Psi = b^i F_i(\sigma) \ .
\eeq

In summary, we start with the minimal extended action, and add
`trivial' (usually quadratic) terms with non-minimal fields (or
antifields).
\beq
S_{tot} = S_{min} +S_{nm}
\eeq
This is the extended action in {\em classical basis}. Then one
performs a canonical transformation on the fields and antifields to
go to the {\em gauge-fixed basis}. By definition, in such a basis, the
part of the action without antifields, denoted as the {\em gauge-fixed
action}
\beq
S_Q= S_{tot} |_{\Phi_A^* =0}\ ,
\eeq
has invertible kinetic terms. Note that, if we are only
interested in the gauge-fixed action and not in its BRST
transformations, the canonical transformation amounts to the
substitution
\beq
\Phi_A^* \rightarrow \tilde\Phi_A^* + \frac{\d \Psi}{ \d \Phi^A} \ .
\eeq
However, after the canonical transformation, we omit the tildes.

This summarizes the BV quantization procedure. For justification and
further details one should consult a standard reference.
Suitable short reviews can be found in \cite{bvshortrev}. For longer
accounts about BV in a Lagrangian setting, see \cite{GomisParis,BVboek}.

It is clear that gauge-fixing is not a straightforward procedure.
Different choices of non-minimal fields and gauge fermions lead to
different propagators. It is a priori not guaranteed that a theory
can be gauge-fixed without breaking locality and covariance.

We will see that in our specific system one encounters many features of the
BV procedure, which will be explained in a case-by-case manner.
However, the only principles that have to be kept in mind are those
which were explained in this section.

\section{First steps}
\subsection{The minimal extended action} \label{minextact}
{\em Henceforth, for simplicity of notation, we omit the integration
symbols and an overall minus sign}, $-\int d^4x d^4\th$, in the
definition of the action. Thus, the classical action \eqn{Scl} is
written as
\begin{equation}
S_{cl}= \bar
\sigma^{\dot \alpha}\bar D_{\dot \alpha} D_\alpha\sigma^\alpha\ .
\label{Scl0}
\end{equation}
The quantization of this classical action starts by
defining the minimal extended action, \eqn{Smin}. The
minimal\footnote{It is always possible to introduce extra zero modes,
as we shall do later on.} zero
modes are defined by symmetric multispinors
$\sigma^{(\alpha\beta)}$,
$\sigma^{(\alpha\beta\gamma)}$, ...~.
\begin{eqnarray}
\delta \sigma^\alpha&=& D_\beta \sigma^{(\beta\alpha)}\nonumber\\
\delta \sigma^{(\beta\alpha)}  &=& D_\gamma \sigma^{(\gamma\beta\alpha)}
\nonumber\\
\delta \sigma^{(\gamma\beta\alpha)}  &=& D_\delta
\sigma^{(\delta\gamma\beta\alpha)}
\nonumber\\
.... &=& ...\ .\label{gaugetr}
\end{eqnarray}
Therefore, we have one part of the minimal extended action for these
transformations
(using now for the ghosts the names used above for the
parameters)
\begin{equation}
S_L=\sum^{\infty}_{i=1}\sigma_{A_i}^* D_\beta \sigma^{(\beta A_i)}
=\sigma^*_\alpha D_\beta \sigma^{(\beta\alpha)}+
\sigma^*_{\beta\alpha}D_\gamma\sigma^{(\gamma\beta\alpha)}+\ldots \ ,
\label{SL}\end{equation}
where $A_i$ is an abbreviation that we will often use for the
symmetrized set of indices $(\alpha_1\ldots \alpha_i)$.
The ghost numbers of $( \sigma^{A_i},\sigma^*_{A_i})$ are $(i-1, -i)$,
and all $\sigma$ are fermionic, all $ \sigma^*$ are bosonic. (See
appendix~\ref{app:conventions} for some remarks on the statistics).

Obviously we have the corresponding contribution for the complex conjugates
with dotted indices, defining $S_R$:
\begin{eqnarray}
S_R&=& \overline{S_L}=
\bar \sigma^*_{\dot \alpha}\bar  D_{\dot \beta}\bar  \sigma^{(\dot \beta
\dot \alpha)}+ \bar
\sigma^*_{\dot \beta\dot \alpha}\bar D_{\dot \gamma}
\bar \sigma^{(\dot \gamma\dot \beta\dot \alpha)}+\ldots \nonumber\\
&=& \bar  \sigma^{(\dot \beta \dot \alpha)}\bar
D_{\dot \beta}\bar \sigma^*_{\dot \alpha} +
\bar \sigma^{(\dot \gamma\dot \beta\dot \alpha)}\bar D_{\dot \gamma} \bar
\sigma^*_{\dot \beta\dot \alpha}+\ldots \ .
\end{eqnarray}
The last line is written (after integration by parts)
because it is easier to consider the terms
in this form. In fact, the fields and antifields can be divided
into either left or right Fields\footnote{We write Fields, with a capital
when it means either a field or an antifield.}. The former will appear only
as the left factors of a term, while right Fields appear only at the
right; see appendix~\ref{app:conventions}. The extended action
$S_{min}= S_{cl}+S_L+S_R$ satisfies the master equation.
\subsection{First level gauge fixing} \label{ss:1stlevel}
The non--minimal
part, a priori arbitrary in the BV approach, should be introduced
with an eye towards  gauge fixing. To complete the kinetic term in
\eqn{Scl0} to an invertible $i\partial_{\alpha\dot \alpha}$, using
\eqn{conv}, we  introduce gauge fixing functions
\beq
F_\a^\ad = D_\a \sigmab^\ad ~~~~,~~~~~\bar{F}_\ad^\a = \Db_\ad \s^\a
\eeq
and corresponding (antighost) fields
$b_\alpha^{\dot \alpha}$ and their complex conjugates $\bar b_{\dot
\alpha}^\alpha$ (both fermions with ghost number $-1$). Note that
the antighosts $b_\alpha^{\dot \alpha}$ have four field components,
while there were only three gauge symmetries described by
$\sigma^{(\alpha\beta)}$.
We will have to compensate for that later on. Indicating
the antifields of the antighosts (bosons of ghost number~0) respectively by
$b^{*\alpha}_{\dot \alpha}$ and $\bar b^{*\dot \alpha}_\alpha$, we
add  to $S_{min}$ the non--minimal term
\begin{equation}
S_{nm,1}= \bar b^{*\dot
\alpha}_\alpha b^{*\alpha}_{\dot \alpha}\ .   \label{Snm1}
\end{equation}
We can then perform a canonical transformation generated by the gauge fermion
\begin{equation}
\Psi_1= b_\alpha^{\dot \alpha} \bar D_{\dot
\alpha}\sigma^\alpha +\bar
\sigma^{\dot \alpha}D_\alpha\bar b_{\dot \alpha}^\alpha \ . \label{Psi1}
\end{equation}
This implies the substitutions
\begin{eqnarray}
b^{*\alpha}_{\dot \alpha} \rightarrow
b^{*\alpha}_{\dot \alpha} +\bar D_{\dot \alpha}
\sigma^\alpha\nonumber\\
\bar b^{*\dot \alpha}_\alpha \rightarrow
\bar b^{*\dot \alpha}_\alpha + \bar \sigma^{\dot \alpha}D_\alpha
\nonumber\\
\sigma^*_\alpha \rightarrow \sigma^*_\alpha +
 b_\alpha^{\dot \alpha}\bar D_{\dot \alpha}\nonumber\\
\sigma^*_{\dot \alpha} \rightarrow \sigma^*_{\dot \alpha}+ D_\alpha
\bar b_{\dot \alpha}^\alpha \ .
\end{eqnarray}
We keep the superspace derivatives to the right of the 'left'
antifields. They act then on unwritten delta functions arising from
the functional differentiation of the gauge fermion. In this way we
never have to interchange superspace derivatives with fields.
We get from the first
terms in $S_L$ and $S_R$ and from the non--minimal action
\begin{eqnarray}
S_{L,1}&\rightarrow & S_{L,1} + b_\alpha^{\dot \alpha}\bar D_{\dot
\alpha}  D_\beta \sigma^{(\beta\alpha)}
\nonumber\\
S_{R,1}&\rightarrow & S_{R,1} + \bar b_{\dot
\alpha}^\alpha D_\alpha  \bar D_{\dot \beta}\bar \sigma^ {(\dot \beta\dot
\alpha)} \nonumber\\
S_{nm,1}&\rightarrow & S_{nm,1} +\bar b^{*\dot
\alpha}_\alpha \bar D_{\dot \alpha} \sigma^\alpha +  \bar
\sigma^{\dot \alpha} D_\alpha b^{*\alpha}_{\dot \alpha} +  \bar
\sigma^{\dot \alpha} D_\alpha \bar D_{\dot \alpha} \sigma^\alpha \ .
\label{nmact1}
\end{eqnarray}
The last term combines with $S_{cl}$
to lead to a good propagator.
The full action at this point is
\begin{eqnarray}
S_1&=&S_{cl}+S_L+S_R+S_{nm,1}= S_{Q,1}+S_{0,1}+S_{*,1}+S_L+S_R \nonumber\\
S_{Q,1}&=& \bar \sigma^{\dot
\alpha}i\partial_{\dot \alpha\alpha} \sigma^\alpha\nonumber\\
S_{0,1}&=&
 b_\alpha^{\dot \alpha}\bar D_{\dot \alpha} D_\beta
\sigma^{(\beta\alpha)} + h.c.
\nonumber\\
S_{*,1}&=&  \bar b^{*\dot \alpha}_\alpha b^{*\alpha}_{\dot \alpha}+
\bar b^{*\dot \alpha}_\alpha \bar D_{\dot \alpha}  \sigma^\alpha +
 \bar \sigma^{\dot \alpha} D_\alpha b^{*\alpha}_{\dot \alpha} \ .
\label{Slevel1}\end{eqnarray}
In order to have eventually kinetic terms which do not mix the
several generations of ghosts, it is most convenient to eliminate
already at this stage the off-diagonal terms in the last line
between $\sigma^\alpha$ and $\bar b^{*\dot \alpha}_\alpha $. Such a
diagonalization can be performed by a redefinition of $\sigma^\alpha$.
To make such redefinitions while keeping the canonical conjugacy of the
fields and antifields, one has to embed them in a canonical
transformation \eqn{Fcan}. In this case we use the generating fermion
\begin{equation}
F(\Phi , \tilde\Phi^*) =\Phi^A \tilde\Phi^*_A -
\tilde{\bar b}{}^{*\dot \alpha}_\alpha \bar D_{\dot \alpha}\frac{1}{\bbox}
i\partial^{\alpha\dot \beta}\tilde{\bar \sigma}^*_{\dot \beta}-
\tilde{\sigma}^*_{ \beta}\frac{1}{\bbox}
i\partial^{\beta\dot\alpha} D_ \alpha\tilde{ b}^{* \alpha}_{\dot\alpha}\ .
\label{Fdiagsigma}
\end{equation}
This leads to
\begin{equation}
S_{*,1} = \bar b^{*\dot \alpha}_\alpha b^{*\alpha}_{\dot \alpha}+
 \bar b^{*\dot \beta}_\alpha  \bar D_{\dot \beta}
\frac{1}{\bbox}i\partial^{\alpha\dot \alpha}D_\beta
b^{*\beta}_{\dot \alpha}+\ \mbox{terms with}\ \sigma^*_\alpha
\mbox{ or }\bar \sigma^*_{\dot \alpha}\ .    \label{diag1}
\end{equation}
The $\sigma^*$ terms are not important for what follows.

Now we have to consider the ghost propagators. Again in $S_{0,1}$ we
have to get terms with the other order of the spinor derivatives,
but moreover there is the problem that there are 4 antighosts and
only 3 ghosts. To obtain an invertible operator, we will thus have to
introduce a fourth ghost. We
just pretend that there is an extra symmetry in the classical action,
whose ghost will be
$\sigma^{[\alpha\beta]}$. The fact that it does not appear in the
transformation of $\sigma^\alpha$ then just implies that there is
an extra zero mode of
the transformation. Thus we write, instead of the second line of
\eqn{gaugetr}, ($C^{\beta\alpha}$
is $i$ times the Levi-Civita symbol $\epsilon^{\beta\alpha}$)
\begin{equation}
\delta\sigma^{\beta\alpha}=D_\gamma\sigma^{(\gamma\beta\alpha)}
+C^{\beta\alpha} \lambda\ ,    \label{zerom1}
\end{equation}
with unsymmetrized $\sigma^{\beta\alpha}$.
Obviously $\lambda$ can be used to gauge away algebraically
$\sigma^{[\alpha\beta]}$ and recover \eqn{gaugetr}. Correspondingly
the field $\lambda$ is a new ghost for ghosts. Therefore the
`minimal extended action' has to be modified. We will do this below.

\section{The full result}    \label{ss:full}
Rather than continuing step by step, we give in this section the final
form of the action. Its structure, as well as that of the gauge
fermion, are encoded in table~\ref{tbl:fields4}.
The arguments which went into the
determination of the set of fields, and the explanation of the steps
leading to the gauge-fixed action, will be given in the following
section.

First, the set of fields, other than the Nielsen-Kallosh
ghosts are given schematically in table~\ref{tbl:fields4} for the
first four levels.
\tabcolsep 1pt
\begin{table}[htb]\caption{Fields up to fourth level.}
\label{tbl:fields4}\begin{center}\begin{tabular}{ccccccccccccccccc}
F &   &   &   &   &   &   & &\ghe{\sigma^{\alpha}}0{-1}&   &   &   &   &   &   &
 &  \\
 &  &   &   &    &   &   &$\swarrow$ & &    &   &   &   &   &   &   &  \\
F &  &      &   &   &   & \ghe{b_{\alpha}^{\dot \alpha}}{-1}0  &   &   &
&\ghe{\sigma^{\alpha_1\alpha_2}}1{-2}&   &
 &   &   &&   \\
 &  &      &   &   & $\swarrow$   &   &   &   & $\swarrow$   &   &   &   &
 &   &  &    \\
F &  &   &   &    \ghe{d^{A_2}_{\dot \alpha}}0{-1}  &
\lre     & \ghe{b_{A_2}^{\dot \alpha}}{-2}1  &
 &   &   &\ghe{\sigma^{A_2\alpha_3}}2{-3}     &   &   &\\
B &  &   &   &    $\nu$  &
\lre     &$ \mu $ &
 &   &   &$\lambda $ &   &   &\\
 &  &   &   $\swarrow$&   &   &   & $\swarrow$   &   &   &   & $\swarrow$
 &   &      & &   & \\
F &  &
\ghe{e_{A_2}^{\dot A_2}}{-1}0 &
&   &   & \ghe{d^{A_3}_{\dot \alpha},d}1{-2}
&  \lre
 & \ghe{b_{A_3}^{\dot \alpha},b}{-3}2  &
 &   &   & \ghe{\sigma^{A_3\alpha_4},\varsigma }3{-4}  &  & \\
B &  &
$ \rho^{\dot \alpha}$ &
&   &   & $\nu^{\alpha}$
&  \lre
 & $\mu_{\alpha}$ &
 &   &   & $\lambda^{\alpha}$  &  & \\
   &$\swarrow$&   &   &   & $\swarrow$   &   &   &   &$\swarrow$    &
 & &   & $\swarrow$ &  &   &     \\
 \ghe{f^{A_3}_{\dot A_2} }0{-1}& \lre   &
\ghe{e_{A_3}^{\dot A_2} }{-2}1 &
&   &   & \ghe{d^{A_4}_{\dot \alpha},d^{\alpha}}2{-3}
&  \lre
 & \ghe{b_{A_4}^{\dot \alpha},b_{\alpha}}{-4}3  &
 &   &   & \ghe{\sigma^{A_4\alpha_5}
 ,\varsigma^{\alpha} }4{-5}      \\
$\tau^{\alpha}_{\dot \alpha}$& \lre   &
$\rho^{\dot \alpha}_{\alpha}$&
&   &   & $\nu^{\alpha_1\alpha_2}$
&  \lre
 &$ \mu_{\alpha_1\alpha_2}$  &
 &   &   & $\lambda^{\alpha_1\alpha_2}$     \\
\end{tabular}\end{center}\end{table}     \tabcolsep 6pt
We indicate for all fields the
(ghost number, ghost number of antifield), and at the left whether
these fields are bosonic or fermionic. For all these fields there are
corresponding complex conjugates. The arrows are related to the
construction of the action, and will be explained later.
We note that entries
in the same row have the same number of field components. For
example, for
level 1 at the right appears the ghost $\sigma^{\alpha_1\alpha_2}$,
{\em not symmetrized}, as discussed above,
in contradistinction to the original
$\sigma^{A_2}\equiv \sigma^{(\alpha_1\alpha_2)}$. Hence, this ghost
contains 4 components as does $b_{\alpha}^{\dot \alpha}$. Similarly,
at the next level the index structure $A_2\alpha_3$ indicates 6
components, as $A_2$ implies a symmetrization in
$(\alpha_1\alpha_2)$, but there is no symmetrization with $\alpha_3$.

Also, one can see in the table that for each field there is a partner
with which it will appear multiplied in the gauge-fixed action. Indeed, a
field of ghost number different from zero has in the table an adjacent
field of opposite ghost number ({\em not} connected by arrows).
See as a first example the first generation ghost and antighost.
Those of ghost number zero will appear multiplied by their own complex
conjugate. Here the classical field is the first example.

All these features are well known, e.g. from the pyramids one obtains
for the quantization of the antisymmetric tensor \cite{AT},and
also the infinite ones for the
Green-Schwarz superstring \cite{Kalloshinf,GSST} or
superparticle \cite{superparticle}. The difference is however
that in those cases, any entry on the same row has the same Lorentz indices.
Here the Lorentz representations vary from left to right, but the
total number of components still matches.

As mentioned at the end of section~\ref{ss:1stlevel} we need extra
ghosts and `fake' gauge invariances expressing the fact that these
ghosts do not appear in the original minimal extended action. We then have
\begin{eqnarray}
S_{min}&=&S_{cl}+S_L +S_R + S_{f,L}+S_{f,R}\nonumber\\
S_{f,L}&=&\sigma^*_{\alpha_1\alpha_2}C^{\alpha_1\alpha_2}\lambda+
\sigma^*_{A_2\alpha_3} C^{\alpha_2\alpha_3}\lambda^{\alpha_1}\nonumber\\
&& +\sigma^*_{A_3\alpha_4} C^{\alpha_3\alpha_4}\lambda^{A_2}
+\varsigma^* C_{\alpha_1\alpha_2}\lambda^{\alpha_1\alpha_2}
+ \ldots \ ,  \label{Sminf}
\end{eqnarray}
where $S_{f,R}$ is the complex conjugate of $S_{f,L}$. An alternative
way to interpret these additions is to say that these are non-minimal
terms, introducing in a trivial way (as auxiliary fields) the fields
$\sigma^{\alpha_1\alpha_2}C_{\alpha_1\alpha_2}$, $\lambda$,
$\sigma^{A_2\alpha_3} C_{\alpha_2\alpha_3}$, ...~, and their
antifields.

To obtain the gauge-fixed action the following choices are made.
The non-minimal terms for the Fields which are not at the extreme
right of each row in the table, are introduced as terms quadratic in the
{\em antifields}. The antifields of {\em ghost number different from
zero} are
multiplied by the ones in the same row to which they are connected
by the double arrow $\stackrel{*}{\longleftrightarrow}$.
Those of ghost number zero
are introduced multiplied by their complex conjugates.
We thus have, to be added to \eqn{Snm1}
\begin{eqnarray}
S_{nm,2}&=&
d^{*\dot \alpha}_{A_2}b^{*A_2}_{\dot \alpha}+
\nu^*\mu^* + h.c.\nonumber\\
S_{nm,3}&=& \bar e^{*\dot A_2}_{A_2}
 e_{\dot A_2 }^{*A_2} +
 \left( d^{*\dot \alpha}_{A_3}b^{*A_3}
 _{\dot \alpha}+\nu^*_\alpha \mu^{*\alpha}+d^*b^*+h.c.\right)
 \nonumber\\
 &&-\bar \rho^*_\alpha i\partial^{\alpha\dot \alpha} \rho^*_{\dot \alpha}
 -\bar \rho'_\alpha i\partial^{\alpha\dot \alpha}\rho'_{\dot \alpha}
\label{S3star}\end{eqnarray}
The last line needs some extra comments. The previous antifields {\em of
ghost number~0} could be introduced in $S_{nm}$ as pure `auxiliary
fields'. This was due to the fact that they had an equal number of
dotted and undotted indices, such that these indices match with
those of the complex conjugates. For the fields where this is not
the case, of which $\rho^{\dot \alpha}$ is the first example, we
have to insert a derivative. Therefore to cancel the extra
propagating mode, we have to introduce a `Nielsen-Kallosh ghost'
\cite{NKgh}, here $\rho'_{\dot \alpha}$. This is, formally, a field
of statistics
opposite to that of the antifield. Thus our Nielsen-Kallosh ghosts
will be formally bosons, to compensate for the fermionic non-minimal
antifields. We will return in section~\ref{ss:NK} to a discussion of
the Nielsen-Kallosh ghosts and their statistics .

The gauge fermion contains terms corresponding to the diagonal arrows
in the table (observe that the ghost numbers of the connected fields
add up to $-1$). Continuing after \eqn{Psi1} for the first level, we add
\begin{equation}
\Psi_2=b_\alpha^{\dot \alpha}
D_\beta d^{(\beta\alpha)}_{\dot \alpha}+b_\alpha^{\dot \alpha}
i\partial^\alpha{}_{\dot
\alpha}\nu
+b_{A_2}^{\dot \alpha}\bar D_{\dot
\alpha}\sigma^{A_2}
+\ft12\mu C_{\alpha\beta}\sigma^{\beta\alpha} +h.c.
\label{Psi2}
\end{equation}
Signs and factors are chosen to obtain simple kinetic terms.
Continuing, we have
\begin{eqnarray}
\Psi_3&=&b^{\dot \alpha}_{A_3}\bar D_{\dot \alpha}
\sigma^{A_3}+\ft23\mu_{\alpha_1} C_{\alpha_2\alpha_3}
\sigma^{A_2\alpha_3}+b^{\dot \alpha}_{A_2}\left(
D_{\alpha_3} d_{\dot \alpha}^{A_3} +i\partial_{\dot \alpha}
^{\alpha_1} \nu^{\alpha_2}+ i\partial^{\alpha_1}_{\dot \alpha}
D^{\alpha_2} d\right) \nonumber\\
&& +e_{A_2}^{\dot A_2}\bar D_{\dot \alpha_1}
d_{\dot \alpha_2}^{A_2}
-2 \rho^{\dot \alpha}\bar D_{\dot \alpha}\nu   + h.c.
\ . \end{eqnarray}
After the corresponding canonical transformations and
various diagonalizations, we find then the following kinetic
terms:
\begin{eqnarray}
S_{Q,2}&=&   b_\alpha^{\dot \alpha}i \partial_{\dot
\alpha \beta} \sigma^{\beta\alpha} +  h.c.
\label{SQ23}\\
S_{Q,3}&=& \bar d^{(\dot \alpha\dot \beta)}_\alpha
i\partial_{\beta\dot \beta}d_{\dot \alpha}^{(\alpha\beta)}
-4\bar \nu \bbox \nu +\left[-\mu \lambda + b^{\dot \alpha}_{(\alpha\beta)}
i\partial_{\gamma\dot\alpha} \sigma^{(\alpha\beta)\gamma} +h.c.\right]
-\bar \rho'_\alpha i\partial^{\alpha\dot \alpha}\rho'_{\dot \alpha}
\ .               \nonumber
\end{eqnarray}
The last term describes the Nielsen-Kallosh ghost. These are the
first terms of the final gauge-fixed action. The invertibility of all
the kinetic operators will be shown in section~\ref{ss:genprop}.

One uses the following terminology. The fields along the upper right
diagonal are called `minimal fields'. They are not completely
minimal, as in fact one could do with only the symmetric spinors
$\sigma^{(A_i)}$. The others, introduced in \eqn{Sminf} to obtain
suitable kinetic terms, are called `catalysts' \cite{superspace}
(sec. 7.3.c). The `classical fields' are just the $\sigma^\alpha$ at
the top. The minimal ones in the second row are `ghosts'. All the other
minimal fields are called `ghosts for ghosts'. The next diagonal
down and to the right are the antighosts, followed by the 'hidden
ghosts' \cite{hiddengh}. However, in the BV formalism all these
Fields are treated in the same way, and we will in what follows just
denote all of them as `ghosts'.

\section{The techniques}
\subsection{The introduction of non-minimal fields}
\label{ss:intronm}
We will now give the arguments that enter in the determination of the
fields which occur in table~\ref{tbl:fields4}. It will be useful to
present them also in a schematic way in table~\ref{tbl:fieldsconj}.
In that table all the fields are split into their irreducible
Lorentz representations. For instance, the tensor
$\sigma^{\alpha_1\alpha_2}$ at the first ghost level in
table~\ref{tbl:fields4} contains a symmetric and an antisymmetric part.
The antisymmetric part is equivalent to a scalar. Therefore we
represent this field as \ffm{20}{00}, where the first column
symbolizes the symmetric part, and the second column the
antisymmetric part. In such a column, the
upper/lower number is the number of symmetrized undotted/dotted
indices in that representation.
\tabcolsep 1pt
\begin{table}[htb]\caption{Fields up to sixth level.}
\label{tbl:fieldsconj}\begin{center}\begin{tabular}{ccccccccccccc}
&&&&&&\ff{1}{0}&&&&&&\\
&&&&&\ff{1}{1}&&\ff{20}{00}&&&&&            \\
&&&&\ff{2}1;\ff 00&\ff *\leftrightarrow  &\ff{2}1;\ff 00&
&\ff{31}{00};\ff 00 &&&&\\
&&&\ff22;\ff10&&\ff{30}{10};\ff10&\ff *\leftrightarrow
&\ff{30}{10};\ff10& &\ff{420}{000}; \ff10&&&\\
&&\ff32;\ff11 &\ff *\leftrightarrow &\ff32;\ff11  &
&\ff{41}{10};\ff{20}{00}&\ff *\leftrightarrow &
\ff{41}{10};\ff{20}{00}&&   \ff{531}{000};\ff{20}{00}&&\\
&\ff33;\ff21&&\ff{40}{20};\ff21 &\ff *\leftrightarrow &
\ff{40}{20};\ff21 & &\ff{520}{100};\ff{31}{00}&\ff *\leftrightarrow &
\ff{520}{100};\ff{31}{00}&   &\ff{6420}{0000};\ff{31}{00}&\\
\ff43;\ff22&\ff *\leftrightarrow & \ff43;\ff22 &&\ff{51}{20};\ff{30}
{10}&\ff *\leftrightarrow & \ff{51}{20};\ff{30}
{10}&&\ff{631}{100};\ff{420}{000}& \ff *\leftrightarrow &
\ff{631}{100};\ff{420}{000}&&\ff
{7531}{0000};\ff{420}{000}
\end{tabular}\end{center}\end{table}     \tabcolsep 6pt

The classical fields and the minimal ghosts introduced in
section~\ref{minextact} are the columns \ffm n0 in the entry
at the right of the $n$-th row.
To introduce the other fields, the following types of arguments are
used:
\begin{enumerate}
\item \label{type1} Non-minimal Fields can be introduced to obtain
appropriate gauge fixing for the fields one level
higher. A non-minimal Field enters in the construction of the
quadratic term involving the field to which it is connected by
a diagonal arrow in table~\ref{tbl:fields4}.
\item \label{type2} All entries on the same horizontal line should have the same
number of components. For those {\em not} connected by arrows this is
because they get multiplied to each other in the kinetic terms in
the gauge-fixed basis, and these kinetic operators should be
invertible.
\item \label{type3} The ones which are connected with
$\stackrel{*}{\longleftrightarrow}$ should have the same index
structure in order that we can build a non--minimal extended action
in the classical basis by multiplying their antifields.
\item \label{type4} If ghosts are introduced in the right diagonal
of minimal Fields, then they should be compensated by ghosts for
these ghosts in the next ghost generation.
\end{enumerate}

Let us now see how this goes in practice.
Typically after a certain level of gauge fixing, ghosts at the next
level occur with a kinetic term involving $\bar D_{\dot \alpha}
D_\beta$, see e.g. after the first level, $b_\alpha^{\dot \alpha}
\bar D_{\dot \alpha}D_\beta\sigma^{(\beta\alpha)}$ in $S_{0,1}$ of
\eqn{Slevel1}.
We want to add a term with $D_\beta\bar D_{\dot \alpha}$ to
complete the previous one to a space-time derivative. This we can
obtain from a square of non-minimal antifields with index structures
with one more undotted index for the left one and
one more dotted index for the right one. In our example this is the
reason for introducing the first term $d^{*\dot \alpha}_{(\alpha\beta)}
b^{*(\alpha\beta)}_{\dot \alpha}$
in $S_{nm,2}$ in \eqn{S3star}. Subsequently one uses
the shift of these antifields as dictated by
the first and third terms in \eqn{Psi2} to replace
$d^{*\dot \alpha}_{(\alpha\beta)}$ by
$b_{(\alpha}^{\dot \alpha}D_{\beta)}$ and
$b^{*(\alpha\beta)}_{\dot \alpha}$ by
$\bar D_{\dot \alpha}\sigma^{(\alpha\beta)}$. This is thus an
argument of type~\ref{type1}.

In the example we would thus obtain the kinetic term
$b_\alpha^{\dot \alpha} \partial_{\beta
\dot \alpha} \sigma^{(\beta\alpha)}$, which is not invertible
because the number of field components do not match. Therefore we
still need a term $b_\alpha^{\dot \alpha} \partial_{\beta
\dot \alpha} \sigma^{[\beta\alpha]}$. This we can obtain from a term
$\nu^*\mu^*$ in the non-minimal action $S_{nm,2}$, with a gauge
fermion leading to the appropriate shifts of these antifields as
obtained from the second and fourth terms in \eqn{Psi2}. This
argument accounts for all the terms in $\Psi_2$. We have thus seen how the
Lorentz structure of $b_\alpha^{\dot \alpha}$ forces us to relax the
symmetrization of $\sigma^{A_2}$ to $\sigma^{\alpha_1\alpha_2}$.
These are arguments of type~\ref{type2}
for non-minimal Fields.

The introduction of $\sigma^{[\alpha_1\alpha_2]}$ is not done by adding
squares of antifields to the non-minimal action.
As explained at the end of section~\ref{ss:1stlevel} its presence
can be seen as a fake gauge invariance, and leads to
the introduction of ghosts for it, in this case $\lambda$.
Similarly, the structure of
$b_{A_2}^{\dot \alpha}$ forces us to relax $\sigma^{A_3}$ to
$\sigma^{A_2\alpha_3}$. These are thus arguments of type~\ref{type4}
for introducing new Fields.

Taking the argument of type~\ref{type3} also into account, one can
now reconstruct the tables. Let us start by following the
{\em left downward} diagonal in table~\ref{tbl:fieldsconj}, and first for
the fermionic fields (those before the semicolon). The argument of
type~\ref{type1} leads in an alternating manner to the addition of a
dotted or of
an undotted index, which explains that representation. Moving in a row
to the right, we either have the argument of type~\ref{type3}, that
the index structure should be the same, or of type 2, pairing first a
representation \ffm nn to \ffm{(n+1)0}{(n-1)0}, and in general,
pairings of ghosts
\begin{equation}
\mbox{\ffm m n} \ \sim \ \mbox{\ffm{(m+1)\,m}{(n-1)\,0}}\ .
\label{genkinform}
\end{equation}
It is this kind of general kinetic term which we will encounter and
treat in section~\ref{ss:genprop}.
Following the horizontal row to the right, we end up with ghost
structure as e.g. in the fourth line \ffm{420}{000}. This has in
comparison to the minimal ghost \ffm 40, the extra ghosts of structure
\ffm{20}{00}. For these, the argument of type~\ref{type4} leads to the
introduction of the bosonic ghosts, after the semicolon on the next
ghost level. In this particular case it explains the \ffm{20}{00} after the
semicolon of the extreme right entry in the fifth line. And in
general, therefore, the index structure of these bosonic ghosts in
the extreme right entry are thus the same as the fermionic ones in
the previous generation apart from the leading one.

For the bosonic ghosts in the other entries we can then apply the same
reasoning as above for the fermionic ones, which completes the
arguments leading to table~\ref{tbl:fieldsconj}.

\subsection{Nielsen-Kallosh ghosts and their statistics.}  \label{ss:NK}

As mentioned in section~\ref{ss:full}, in some cases we have
to introduce the non-minimal antifields in the extended action with a
derivative, and therefore have to cancel a contribution to possible loops,
by adding Nielsen-Kallosh ghosts \cite{NKgh}, e.g.
\begin{equation}
\bar \rho'_\alpha i\partial^{\alpha\dot \alpha}
\rho'_{\dot \alpha}\ .\label{NKform}
\end{equation}
This occurs when these
antifields have ghost number zero and an unbalanced number of dotted
and undotted spinor indices. One can easily check from the
table~\ref{tbl:fieldsconj} that this happens only for the bosonic
ghosts (with fermionic antifields) at every other level starting with
$\rho^*_{\dot \alpha}$.

There is a subtlety here. Interpreting $\rho'_{\dot \alpha}$ as a bosonic field,
its contribution to the action, as written in \eqn{NKform},
is purely imaginary (and adding the hermitian conjugate leads thus to a zero
result, or more
exactly, to a total derivative).
The same problem occurs when one introduces Pauli-Villars \cite{PV,BVboek}
regulator fields.
These should also have statistics opposite
to those of the fields whose divergent loop contributions are regulated.
However, using the kinetic term $\phi\bbox\phi$ for
fermionic fields gives again a total derivative.
Here the correct way is to replace the boson $\rho'_{\dot \alpha}$ by
two bosons $\rho'_{\dot \alpha}$ and $\rho''_{\dot \alpha}$
and a fermion $\rho'''_{\dot \alpha}$ (realistic solution)\footnote{The
terminology is borrowed from the paper by Pauli and Villars, who
used it in a somewhat different sense however.}. The action for these fields
is then
\begin{equation}
\bar \rho'_\alpha i\partial^{\alpha\dot \alpha} \rho''_{\dot \alpha}
+\rho'_{\dot \alpha} i\partial^{\alpha\dot \alpha}\bar  \rho''_{ \alpha}
+\bar \rho'''_\alpha i\partial^{\alpha\dot \alpha} \rho'''_{\dot \alpha}\ ,
\label{NKreal}\end{equation}
which can be used without any problems in a
path integral. The effect is the same as what one obtains
from \eqn{NKform} (with fermionic $\rho'$),
adding a minus sign by hand in all loops (a formalistic solution).

\subsection{General form of propagators} \label{ss:genprop}

In general we obtain kinetic terms which are formally of the
type \eqn{genkinform}. Explicitly, they take the following form:
\begin{equation}
S_k=\Phi_{B_m}^{\dot A_n}\left( i\partial_{ \beta_{m+1}\dot\alpha_n}
\Lambda^{B_{m+1}}_{\dot A_{n-1}}+i\partial^{\beta_1}_{\dot
\alpha_1}\cdots i\partial^{\beta_n}_{\dot \alpha_n}\Gamma^{\beta_{n+1}
\cdots \beta_m}\right)\ ,  \label{generickin}
\end{equation}
with $m\geq n-1$. If $m=n-1$, there is no $\Gamma$ term.
A simple case is obtained for $m=1$, $n=1$, when
one may write $\Gamma$ as the antisymmetric part of
$\Lambda^{\alpha\beta}$:
\begin{equation}
\Gamma =\ft12 C_{\alpha\beta}\Lambda^{\beta\alpha}\ ,
\end{equation}
such that the two terms combine to
\begin{equation}
\Phi^{\dot \alpha}_{\beta_1}i\partial_{\beta_2\dot
\alpha}\Lambda^{\beta_1\beta_2}\ .
\end{equation}
This is how these terms appear in $S_{Q,2}$, and similarly in $S_{Q,3}$
in \eqn{SQ23}. At higher levels such a simple rewriting is not
always possible (e.g. with $e^{\dot A_2}_{A_2}$ as  $\Phi$, and
$d^{A_3}_{\dot \alpha}$, $d$ as $\Lambda$,
$\Gamma$ with $m=n=2$).

$S_k$ is an
invertible kinetic term, as we will show by obtaining explicitly
the propagator, as the inverse of the kinetic operator.
It can be defined as the tensors $P_1$ and $P_2$, where
\begin{equation} \left[
i\partial_{\beta_{m+1}(\dot\alpha_n}P_1{}^{B_{m+1}}_{\dot A_{n-1})}
{}^{;\dot D_n}_{;C_m}+
i\partial^{(\beta_1}_{\dot \alpha_1}\cdots
i\partial^{\beta_n}_{\dot \alpha_n}
P_2{}^{\beta_{n+1}
\cdots \beta_m)}{}^{;\dot D_n}_{;C_m}\right]E_{\dot D_n}^{C_m}
=E^{B_m}_{\dot A_n}\ ,
\end{equation}
for an arbitrary (symmetric) tensor $E$. The tensor $P_1$ describes
then the propagator between the $\Phi$ and $\Lambda$ field, while
$P_2$ describes the propagator between $\Phi$ and $\Gamma$. The
solution for these tensors is given by
\begin{eqnarray}
\lefteqn{-P_1{}^{B_{m+1}}_{\dot A_{n-1}}
{}^{;\dot D_n}_{; C_m}E^{C_m}_{\dot D_n}=n
\frac{1}{\bbox} i\partial^{\dot \gamma(\beta_{m+1}}E^{B_m)}_{
\dot A_{n-1}\dot \gamma}}\nonumber\\ &&+{n\choose 2}  \frac{1}{\bbox^2}
i\partial^{\dot \gamma_1(\beta_1}i\partial^{\dot \gamma_2\beta_2}
i\partial_{\delta(\dot \alpha_1}E^{\beta_3\cdots\beta_{m+1})\delta}_{
\dot\alpha_2\dot \alpha_{n-1})\dot \gamma_1\dot \gamma_2 }+\ldots\nonumber\\
&&
 +{n\choose  n}\frac{1}{\bbox^n}
i\partial^{\dot \gamma_1(\beta_1}\cdots i\partial^{\dot \gamma_n\beta_n}
i\partial_{\delta_1\dot \alpha_1}\cdots i\partial_{\delta_{n-1}\dot \alpha_{n-1}}
E^{\beta_{n+1}\cdots\beta_{m+1})\delta_1\cdots\delta_{n-1} }_{
\dot \gamma_1\cdots \dot\gamma_{n} }
\nonumber\\
\lefteqn{ -P_2^{\beta_{n+1}\cdots\beta_m }{}^{;\dot D_n}_{;C_m}E^{C_m}
_{\dot D_n} = -\frac{m+1-n}{m+1}\,\frac{1}{\bbox^n}\,
i\partial_{\beta_1}^{\dot \delta_1}\cdots i\partial_{\beta_n}^{\dot \delta_n}
E^{B_m}_{ \dot D_n}}\label{genpropagator}
\end{eqnarray}
This shows the invertibility of the kinetic terms.

\subsection{Diagonalizations}

We have already given one example of diagonalization of terms in
the extended action in section~\ref{ss:1stlevel}. It is important to
observe that one is not allowed to drop the terms with antifields at
intermediate steps. Indeed, since we perform the gauge fixing in several
(infinite number of) steps, the terms depending on antifields whose
fields will appear in later gauge fermions are still relevant.
The final kinetic terms which we get arise often from various
sources. The $\bar \nu\bbox\nu$ term, for instance, in $S_{Q,3}$ in
\eqn{SQ23}, gets first contributions from inserting the substitution
generated by the gauge fermion $\Psi_2$ into the two terms of
\eqn{diag1}, one of which already arose from a previous
diagonalization. The canonical transformation generated by $\Psi_3$
gives rise to other terms. Some of these generate kinetic terms
mixing $\nu$ with $d_{\dot \alpha}^{A_2}$. Subsequently we diagonalize the
kinetic terms of $d_{\dot \alpha}^{A_2}$ in a manner similar to that
for $\sigma^\alpha$ in \eqn{Fdiagsigma}. All these contributions at the
end lead to the $\bar \nu\bbox\nu$ result. This is also the consequence
of some choices of signs and factors in the gauge fermions.

To perform these diagonalizations, one uses the propagator obtained
in \eqn{genpropagator}. Indeed, one can use the following general
procedure: the extended actions which we have to diagonalize are
of the general form
\begin{eqnarray}
S&=&\Phi_{B_m}^{\dot A_n}\left( i\partial_{ \beta_{m+1}\dot\alpha_n}
\Lambda^{B_{m+1}}_{\dot A_{n-1}}+i\partial^{\beta_1}_{\dot
\alpha_1}\cdots i\partial^{\beta_n}_{\dot \alpha_n}\Gamma^{\beta_{n+1}
\cdots \beta_m}\right)\nonumber\\
&&+U_{B_{m+1}}^{\dot A_{n-1}}   \Lambda^{B_{m+1}}_{\dot A_{n-1}}
+V_{A_{m-n}}  \Gamma^{A_{m-n} }
+ \Phi_{A_m}^{\dot A_n}  W^{A_m}_{\dot A_n}
\ ,
\end{eqnarray}
where $U$, $V$ and $W$ are (symmetric) tensors depending on other
Fields. We then perform in principle a canonical transformation
with generating fermion
(if there are antifields in $U$, $V$ and $W$,
one should replace them here with their new (tilde) antifields)
\begin{eqnarray}
F(\Phi , \tilde\Phi^*) &=&\Phi^A \tilde\Phi^*_A
+U_{B_{m+1}}^{\dot A_{n-1}} P_1{}^{B_{m+1}}_{\dot A_{n-1}}
{}^{;\dot D_n}_{; C_m}{\tilde\Phi}{}^{*\,C_m}_{\dot D_n}
+ V_{A_{m-n}}P_2^{A_{m-n} }{}^{;\dot D_n}_{;C_m}{\tilde\Phi}{}^{*\,C_m}
_{\dot D_n}               \nonumber\\ &&
+{\tilde\Lambda}{}_{B_{m+1}}^{*\dot A_{n-1}} P_1{}^{B_{m+1}}_{\dot A_{n-1}}
{}^{;\dot D_n}_{; C_m}W^{C_m}_{\dot D_n}
+ {\tilde\Gamma}^*_{A_{m-n}}P_2^{A_{m-n} }{}^{;\dot D_n}_{;C_m}W^{C_m}
_{\dot D_n}\ .
\end{eqnarray}
That leads then to (all Fields in the new basis)
\begin{eqnarray}
S&=&\Phi_{B_m}^{\dot A_n}\left( i\partial_{ \beta_{m+1}\dot\alpha_n}
\Lambda^{B_{m+1}}_{\dot A_{n-1}}+i\partial^{\beta_1}_{\dot
\alpha_1}\cdots i\partial^{\beta_n}_{\dot \alpha_n}\Gamma^{\beta_{n+1}
\cdots \beta_m}\right)\nonumber\\
&&-U_{B_{m+1}}^{\dot A_{n-1}}  P_1{}^{B_{m+1}}_{\dot A_{n-1}}
{}^{;\dot D_n}_{; C_m} W^{C_m}_{\dot D_n}  -
V_{A_{m-n}} P_2^{A_{m-n} }{}^{;\dot D_n}_{;C_m}W^{C_m}_{\dot D_n}\ .
\end{eqnarray}
This one lemma in sufficient to perform all the diagonalizations.

\section{Conclusions}

In this work we have described the superspace BV quantization
of the complex linear superfield $\S$ which provides an alternative
description of the component $N=1$ scalar multiplet. As mentioned in
the introduction, and due partly to the presence of superspace
derivatives with a single spinor index, the treatment of our system
brings in some features not encountered in previous applications of the
BV method.

As indicated in section 2, at the classical level the complex linear
superfield  (nonminimal scalar multiplet) with superspace lagrangian
$\Sigmab \Sigma$
 differs from a chiral superfield with lagrangian $\Phib \Phi$
 only in their auxiliary field
structure - yet the quantization of the former gives rise to difficulties
which are not present for the latter, and makes any practical applications
rather difficult to envisage. Thus, our quantization reveals more
about the BV approach than about the physics described by the nonminimal
multiplet. In particular, issues such as whether different sets of
auxiliary fields lead to different quantum physics (as they appear to
do in supergravity \cite{superspace}) are not easy to investigate here.

The main problem with the linear superfield is that the gauge transformation
$\d \s^\a = D_\b \s^{(\b \a)}$ ``overshoots'' in the gauge manifold.
We have considered an alternative way of presenting the gauge transformations
in terms of chiral spinor and chiral symmetric bispinor superfields
$\S^\a$, $\S^{(\a \b )}$ as
\beq
\d \s^\a = D^2 \S^\a +D_\b \S^{(\a \b )} = D_\b \left[- \ft{2}{3} D^{(\a}
\S^{\b )}+\Sigma ^{(\alpha\beta)}\right]\ .
\eeq
However, the chirality of the gauge parameters implies that at
the component level some gauge components of $\s^\a$, absent in the action,
 transform with space-time derivatives. This introduces extra
propagating modes in the ghost action, which have to be cancelled in a
manner which may  not be manifestly supersymmetric.
\vspace{5mm}

We have constructed the gauge-fixed action of the complex linear
superfield, which contains an infinite tower of ghosts. The structure
of all the fields that enter can be inferred from the part exhibited
in tables~\ref{tbl:fields4} and~\ref{tbl:fieldsconj}. The full
gauge-fixed action is obtained as follows:
\begin{itemize}
\item The fields of non-zero ghost number have partners adjacent to
them in the table with opposite ghost number. These fields are
multiplied together in the action. The generic
pattern is \eqn{genkinform}, which explicitly takes the form
\eqn{generickin}. We obtained their propagator in
\eqn{genpropagator}.  However, some of them enter as non-propagating
auxiliary fields according to their dimension and index structure.
\item Fields of ghost number zero appear in the action multiplied
with their complex conjugates.
\item Finally there are Nielsen-Kallosh ghosts not indicated in the
table as discussed after \eqn{NKform}.
\end{itemize}

\medskip
\section*{Acknowledgments.}

\noindent
We thank W. Siegel and W. Troost for discussions. This work was
supported by the European Commission TMR programme
ERBFMRX-CT96-0045, in which D.Z. is associated to the University of Torino.
Part of this work was performed during visits of
the authors at CERN and at the Newton Institute in Cambridge, and
visits of M.G. and A.V.P. in Milano University. The authors
gratefully acknowledge the hospitality at all these places.
\newpage

\appendix
\section{Conventions}
 \label{app:conventions}
The superspace conventions we use are
\begin{eqnarray}
D^\beta&=&C^{\beta\gamma}D_\gamma \nonumber\\
C_{\alpha\beta}C^{\gamma\delta}&=&
\delta_\alpha^\gamma\delta_\beta^\delta  -
\delta_\alpha^\delta\delta_\beta^\gamma\ ;\qquad C_\alpha{}^\beta
=\delta_\alpha{}^\beta\nonumber\\
2 A_{[\alpha}B_{\beta]}&=&C_{\beta\alpha}A^\gamma B_\gamma\nonumber\\
{}[\alpha\beta]&=&\ft12(\alpha\beta-\beta\alpha)\nonumber\\
i\partial_{\alpha\dot \alpha}&=&D_\alpha\bar D_{\dot \alpha}+
\bar D_{\dot \alpha}  D_\alpha      \nonumber\\
D^2&=& \ft12 D^\alpha D_\alpha\nonumber\\
\bbox&=&\ft12 \partial^{\alpha\dot \alpha}\partial_{\alpha\dot \alpha}=
D^2\bar D^2+\bar D^2 D^2-D^\alpha\bar  D^2D_\alpha\nonumber\\
D_\alpha D_\beta&=&C_{\beta\alpha}D^2    \label{conv}
\end{eqnarray}

\noindent{\bf Left and right fields}

In the tables, we consider only half of the fields.
There are also the complex
conjugates which form a similar table. We have indicated by a
diagonal arrow the fields which are connected in the
gauge fermion.  One could also assign a helicity to all the fields and
antifields such that in the extended action terms occur in which only
Fields of different helicity are multiplied together.
Any term in the action or in
the gauge fermion is of the form $L\ O\ R$, where
$O$ is a possible superspace operator or spacetime
derivative. We assign helicity $R$ to all Fields with upper dotted
indices and lower undotted indices\footnote{Throughout this work we
do not raise or lower indices on Fields.},
and vice versa for $L$ helicity. Fields and
antifields have opposite helicities, and so do Fields and their
complex conjugates. In the tables the $\sigma$, $\lambda$,
$\varsigma$, $d$, $\nu$, $f$, $\tau$ fields
on the NW-SE diagonals have $R$ helicity. The other diagonals $b$, $\mu$,
$e$, $\rho$ have $L$ helicity.

\noindent{\bf Statistics and superspace rules}

A more delicate point is how one treats the commutation or
anticommutation of fields and the superspace coordinate $\theta^\alpha$
and $D_\alpha$ which are fermionic. From \eqn{components}, $\Sigma$
is bosonic, and therefore $\sigma^\alpha$ is fermionic. Since
antifields have opposite statistics to
fields, $\sigma^*_\alpha$ is bosonic. Then, \eqn{SL}
implies that all $\sigma$-fields with any index structure are
fermionic. The gauge fermion is always fermionic.
It follows that $b^\alpha_{\dot \alpha}$ is fermionic and therefore
$\sigma^{[\alpha\beta]}$ is also fermionic. Its ghost $\lambda$ is
bosonic. Proceeding in this manner one arrives at the assignments in
the table~\ref{tbl:fields4}.

We note that these assignments, which
are consistent with ordinary space BV methods, lead to rules which
are different from those used in other superspace calculations, where
objects with even or odd number of spinor indices are assigned
(ab)normal statistics depending on their ghost number. Consequently
some care is required when transferring spinor derivatives past
Fields. We have proved that both assignments of statistics lead to
the same results. The proof relies on the helicity structure
explained above.

\noindent{\bf Dimensions}

In assigning dimensions to the Fields, there is some freedom of
choice.  A
suitable choice can be as follows: One assigns dimension 1 to a
spacetime derivative $\partial_{\alpha\dot \alpha}$ and $\ft12$ to a
superspace derivative $D_\alpha$. The action has dimension 1, and the
gauge fermion $\Psi$ has dimension $\ft12$. The dimensions of a field
and antifield add up to $\ft12$. All the fermion fields
$\sigma^{A_n\alpha_{n+1}}$,
$b^{\dot \alpha}_{A_n}$, $d_{\dot \alpha}^{A_n}$, $e^{\dot A_2}_{A_n}$,
$f_{\dot A_2}^{A_n}$ have dimension 0,
therefore corresponding antifields have dimension $\ft12$. The boson fields
$\lambda$, $\mu$, $\rho$ have dimension $\ft12$, and their antifields
dimension 0. However $\nu$ has dimension $-\ft12$ and thus
$\nu^*$ has dimension 1. The field $\rho'$ has dimension equal to that of
$\rho^*$, namely 0.


\begin{thebibliography}{999}
\addcontentsline{toc}{chapter}{Bibliography}
\bibitem{superspace} S.J. Gates, M.T. Grisaru, M. Ro\v{c}ek and W. Siegel,
{\em Superspace} (Addison-Wesley, 1983).
\bibitem{WZNW} S.J.Gates, Phys. Lett. {\bf B365} (1996) 132, hep-th/9508153;\\
S.J.Gates, M.T. Grisaru, E.E. Knutt-Wehlau, M. Ro\v{c}ek and O.A. Soloviev,
Phys. Lett. {\bf B} (to be published), hep-th/9612196.
\bibitem{YMSG} M.T. Grisaru, A. Santambrogio and D. Zanon,
Nucl. Phys {\bf B487} (1997) 174, hep-th/9610001,
\bibitem{BV} I.A. Batalin and G.A. Vilkovisky, Phys. Rev. {\bf
D28} (1983) 2567 (E:{\bf D30} (1984) 508).
\bibitem{Deo} B.B. Deo and S.J. Gates, Nucl. Phys. {\bf B254} (1985) 87.
\bibitem{BVcan} I.A. Batalin and G.A. Vilkovisky, Nucl. Phys. {\bf B234}
(1984) 106;\\
 W. Siegel, Int. J. Mod. Phys. {\bf A4} (1989) 3705;\\
 W. Troost, P. van Nieuwenhuizen and A. Van Proeyen,
Nucl. Phys. {\bf B333} (1990) 727.
\bibitem{bvshortrev} A.  Van Proeyen, in Proc.  of the Conference {\it
Strings \& Symmetries 1991}, Stony Brook, May 20--25, 1991, eds.  N.
Berkovits et al., (World Sc.  Publ.  Co., Singapore, 1992), p.
388.;\\
 W. Troost and A. Van Proeyen,
in {\em Strings 93}, proceedings of the
Conference in Berkeley, CA,  24-29 May 1993,
eds. M.B. Halpern, G. Rivlis and A. Sevrin, (World Sc. Publ. Co.,
Singapore), p. 158; hep-th/9307126;\\
 W. Troost and A. Van Proeyen,
in {\em Strings and Symmetries}, Lecture Notes in Physics,
Vol. 447, Springer-Verlag,
eds. G. Aktas, C. Saclioglu, M. Serdaroglu, p. 183,
hep-th/9410162.
\bibitem{NKgh}  N.K. Nielsen, Nucl. Phys. {\bf B140} (1978) 499;\\
 R.E.~Kallosh, Nucl.\ Phys.\ {\bf B141} (1978) 141.
\bibitem{GomisParis}  J. Gomis, J. Par\'{\i}s and S. Samuel,
Phys. Rep. {\bf 259} (1995) 1; hep-th/9412228.
\bibitem{BVboek} W. Troost and A. Van Proeyen, {\em An introduction
to Batalin--Vilkovisky Lagrangian quantization},
Leuven Notes in Math. Theor. Phys., in preparation.
\bibitem{AT} D.Z. Freedman, CalTech report CALT 68--624 (1977),
unpublished;\\
D.Z. Freedman and P.K. Townsend, Nucl. Phys. {\bf B177} (1981) 282;\\
S.P. de Alwis, M.T. Grisaru and L. Mezincescu, Phys. Lett. {\bf B190}
(1987) 122; Nucl. Phys. {\bf B303} (1988) 57;\\
A.H. Diaz, Phys. Lett. {\bf B203} (1988) 408;\\
C. Battle and J. Gomis, Phys. Rev. {\bf D38} (1988) 1169;\\
L. Baulieu, E. Bergshoeff and E. Sezgin, Nucl. Phys. {\bf B307} (1988)
348;\\
G. Barnich, R. Constantinescu and P. Gr\'egoire, Phys. Lett. {\bf B293}
(1992) 353.
\bibitem{Kalloshinf} R.E. Kallosh, JETP Lett. {\bf 45} (1987) 365; Phys.
Lett. {\bf B195} (1987) 369.
\bibitem{GSST} W. Siegel,
in {\it 'Strings 89'}, eds. R. Arnowitt et al., World Scientific, 1990;\\
 U. Lindstr\"om, M. Ro\v{c}ek, W. Siegel, P. van
Nieuwenhuizen and A.E. van de Ven, Nucl. Phys. {\bf B330} (1990) 19.
\bibitem{superparticle} A. Mikovi\'c, M. Ro\v{c}ek, W. Siegel, P. van
Nieuwenhuizen, J. Yamron and A.E. van de Ven, Phys. Lett. {\bf B235} (1990)
106;\\
E.A. Bergshoeff, R. Kallosh and A. Van
Proeyen, Class. Quantum Grav. {\bf 9} (1992) 321.
\bibitem{hiddengh} W. Siegel, Phys. Lett. {\bf 93B} (1980) 170.
\bibitem{PV} W. Pauli and F. Villars, Rev. Mod. Phys. {\bf 21} (1949) 434.
\end{thebibliography}
\end{document}